\documentstyle[preprint,aps]{revtex}
\tighten
\begin{document}
\draft
\preprint{\vbox{Submitted to Physical Review B 
		          \hfill FSU-SCRI-96-1214 \\
                          \null\hfill cond-mat/xxxxxx}}
\title{A Plaquette Basis for the Study of Heisenberg Ladders}
\author{J. Piekarewicz}
\address{Supercomputer Computations Research Institute, \\
         Florida State University, Tallahassee, FL 32306}
\author{J.R. Shepard}
\address{Department of Physics, \\
         University of Colorado, Boulder, CO 80309}
\date{\today}
\maketitle
 
\begin{abstract}
We employ a plaquette basis---generated by coupling the 
four spins in a $2\times2$ lattice to a well-defined 
total angular momentum---for the study of Heisenberg 
ladders with antiferromagnetic coupling. Matrix elements 
of the Hamiltonian in this basis are evaluated using standard 
techniques in angular-momentum (Racah) algebra. We show by 
exact diagonalization of small ($2\times4$ and $2\times6$) 
systems that in excess of 90\% of the ground-state probability 
is contained in a very small number of basis states.  These few 
basis states can be used to define a severely truncated basis
which we use to approximate low-lying exact eigenstates. We show 
how, in this low-energy basis, the isotropic spin-$1/2$ Heisenberg 
ladder can be mapped onto an anisotropic spin-$1$ ladder for which 
the coupling along the rungs is much stronger than the coupling between
the rungs. The mapping thereby generates two distinct energy scales 
which greatly facilitates understanding the dynamics of the original 
spin-$1/2$ ladder. Moreover, we use these insights to define an effective 
low-energy Hamiltonian in accordance to the newly developed 
COntractor REnormalization group (CORE) method. We show how a simple 
range-2 CORE approximation to the effective Hamiltonian to be used 
with our truncated basis reproduces the low-energy spectrum of the 
exact $2\times6$ theory at the $\alt 1\%$~level.
\end{abstract}

\narrowtext

\section{Introduction}
\label{sec:intro}

 Interest in the study of ladder compounds has been stimulated by 
suggestions that these deceptively simple materials could exhibit 
some of the critical behavior believed to be responsible for 
high-temperature (high-$T_{c}$) 
superconductivity~\cite{drs92,dagric96}. Moreover, these 
ladder materials---built one chain at a time---could bridge the
transition from the one-dimensional (1D) chains, where large 
simulations are carried out routinely, to the two-dimensional (2D)
structures that are known to be at the heart of the high-$T_{c}$ 
compounds~\cite{dag94}. Although this might very well be the case, 
the road from one to two dimensions has been full of (quantum) 
surprises~\cite{dagric96}. Indeed, it is now known that some of the 
properties common to the 1D- and 2D-systems---such as the absence of 
a spin gap---are only shared by the odd-leg ladders. In contrast, 
even-leg ladders have a finite spin gap which should manifest itself 
in the form of an exponential behavior of the magnetic susceptibility 
at low temperatures; this (exponential) activation has been 
confirmed experimentally~\cite{azuma94}. Finally, the excitement about
the ladder materials has recently been fueled even further by some 
preliminary reports that suggest a superconducting transition in some 
physical realizations of the ladders at a critical temperature of about 
$T_{c}=12$~K~\cite{uehar96}.

An arsenal of numerical approaches has been employed to elucidate  
the physics of the ladder materials. This includes exact and Lanczos 
diagonalization techniques~\cite{lancz50}, quantum Monte-Carlo 
methods~\cite{binhee92}, and a density-matrix-renormalization-group 
approach~\cite{white92}. While these techniques have achieved a high 
degree of sophistication, the selection of a basis has not. Indeed, 
the overwhelming majority---if not all---of the calculations reported 
in the literature have rely on the traditional ``$S_z$'' basis. 
Undoubtedly, the biggest advantage of the $S_z$ basis is its simplicity. 
Indeed, matrix elements of the Hamiltonian are trivially computed in 
this basis. Yet, it is unlikely that the $S_z$ basis represents the 
``optimal'' choice for the study of Heisenberg antiferromagnets with 
isotropic coupling. 

 Recent discussions of the possible utility of alternative bases 
for the study of the ladder materials by Martins, Dagotto, and 
Riera~\cite{mdr96} have stimulated us to examine this issue. The 
initial step in our program was to rediscover the ``plaquette'' 
basis~\cite{ashmer76} and then to examine how it might be used 
in the study of spin ladders. A particularly appealing feature 
of the plaquette basis---generated from the coupling of the four 
spins in a $2 \times 2$ lattice to a well-defined total angular 
momentum---is that the states in the basis are the eigenstates 
of the largest two-leg ladder that can be solved in closed form. 
By this mere fact, it is clear that much of the important physics 
of the problem has been incorporated into the basis. Obviously, 
an important component of our program is the evaluation of matrix 
elements of the Hamiltonian. It is for this that we rely heavily 
on the sophisticated apparatus of angular-momentum 
algebra~\cite{bieden81,brisat93} that has been developed in 
atomic~\cite{consho51} and nuclear physics~\cite{bohmot69,shafes74} 
over many years. 

 Our paper has been organized as follows. In Sec.~\ref{sec:formal}
we describe the plaquette basis and compute all the necessary
matrix elements of the Hamiltonian. In Sec.~\ref{sec:applic} we
study small $2 \times 4$ and $2 \times 6$ ladders to illustrate
the advantage of our approach. In particular, we define a severely 
truncated basis which includes only the 4 lowest, out of 16, 
1-plaquette energy levels.  We use the newly developed COntractor 
REnormalization (CORE)~\cite{morwei94} method to construct a 
low-energy effective Hamiltonian to be used with the truncated basis 
and find that the exact eigenvalues of the 3-plaquette system are 
very accurately reproduced. Finally, we summarize in Sec.~\ref{sec:concl}.

\section{Formalism}
\label{sec:formal}

 The focus of our paper is the plaquette basis. This basis 
is generated from the coupling of the four spins in a $2 \times 2$ 
plaquette to a well-defined total angular momentum. To define the 
basis and to illustrate most of the techniques it is sufficient to 
concentrate on the $2 \times 4$ (or two-plaquette) antiferromagnet 
with isotropic nearest-neighbor coupling $J\equiv 1$. Moreover, we 
restrict ourselves to the $S=1/2$ case, although the formalism can be 
applied with few modifications to arbitrary spin. Indeed, we exploit 
this flexibility when, in Sec~\ref{sec:applic}, we discuss spin-$1$ 
ladders which are in some sense equivalent to the spin-$1/2$ ladders 
discussed in this section. The model Hamiltonian is given---adopting 
open boundary conditions---by
\begin{equation}
   H=\sum_{\langle i,j \rangle}{\bf S}_{i}\cdot{\bf S}_{j} 
   \equiv H_{0}(1) + H_{0}(2) + V(1,2) \;,
  \label{eightsite}
\end{equation}
where $\langle i,j\rangle$ denotes nearest-neighbor sites and
\begin{mathletters}
\begin{eqnarray}
  H_{0}(1) &\equiv& {\bf S}_{1}\cdot{\bf S}_{2} +
                    {\bf S}_{1}\cdot{\bf S}_{3} +
                    {\bf S}_{2}\cdot{\bf S}_{4} +
                    {\bf S}_{3}\cdot{\bf S}_{4} \;, \\
  H_{0}(2) &\equiv& {\bf S}_{5}\cdot{\bf S}_{6} +
                    {\bf S}_{5}\cdot{\bf S}_{7} +
                    {\bf S}_{6}\cdot{\bf S}_{8} +
                    {\bf S}_{7}\cdot{\bf S}_{8} \;, \\
  V(1,2)   &\equiv& {\bf S}_{3}\cdot{\bf S}_{5} +
                    {\bf S}_{4}\cdot{\bf S}_{6} \;.
\end{eqnarray}
\label{htwoplaquette}
\end{mathletters}
Note that this simple two-leg ladder is formed from two interacting 
$n\!=\!4$ spin chains; spins on the left(right) chain are labeled with 
odd(even) numbers. Moreover, $H_{0}$ represents the Hamiltonian 
of an isolated $2 \times 2$ system, while $V(1,2)$ is the coupling 
``potential''. The $2 \times 2$ Hamiltonian $H_{0}$---a simple
``antiferromagnet''~\cite{ashmer76}---can be re-written in the 
following form:
\begin{equation}
 H_{0}(1) = ({\bf S}_{1} + {\bf S}_{4})   \cdot
            ({\bf S}_{2} + {\bf S}_{3})   =
            {\bf L}_{1} \cdot {\bf L}_{2} =
	    {1 \over 2} 
	    \left({\bf J}_{1}^{2}-{\bf L}_{1}^{2}-{\bf L}_{2}^{2} \right) \;,
 \label{plbasis}
\end{equation}
where we have defined link and plaquette angular-momentum variables,
respectively, as
\begin{equation}
 {\bf L}_{1} \equiv  {\bf S}_{1} + {\bf S}_{4} \;; \quad
 {\bf L}_{2} \equiv  {\bf S}_{2} + {\bf S}_{3} \;; \quad
 {\bf J}_{1} \equiv  {\bf L}_{1} + {\bf L}_{2} =
                     {\bf S}_{1} + {\bf S}_{2} +
                     {\bf S}_{3} + {\bf S}_{4} \;.
 \label{landj}
\end{equation}
The physically appealing feature of the plaquette basis is that 
$H_{0}$ is diagonal in this basis. That is,
\begin{equation}
  H_{0}(1)|l_{1}l_{2},j_{1}m_{1}\rangle = {1 \over 2}
  \Big[j_{1}(j_{1}+1)-l_{1}(l_{1}+1)-l_{2}(l_{2}+1)\Big]
  |l_{1}l_{2},j_{1}m_{1}\rangle \;. 
 \label{etwosite}
\end{equation}
It is interesting to note that one needs to couple the spins along 
the diagonal---which are the only spins that do not interact---in 
order to bring the Hamiltonian into a diagonal form. The eigenvalues
and eigenvectors of the $2 \times 2$ Hamiltonian have been listed in
Table~\ref{tableone}. It is also instructive to write the ground-state
of $H_{0}$, namely, the 
$|l_{1}\!=\!l_{2}\!=\!1,j_{1}\!\!=m_{1\!}=\!0\rangle$ state,
in terms of the $S_{z}$ basis. That is,
\begin{equation}
 |\Psi_{0}\rangle = {1 \over \sqrt{3}} \Big[
                 |\uparrow\downarrow\downarrow\uparrow\rangle +
                 |\downarrow\uparrow\uparrow\downarrow\rangle -
     {1 \over 2} |\uparrow\uparrow\downarrow\downarrow\rangle -
     {1 \over 2} |\uparrow\downarrow\uparrow\downarrow\rangle -
     {1 \over 2} |\downarrow\uparrow\downarrow\uparrow\rangle -
     {1 \over 2} |\downarrow\downarrow\uparrow\uparrow\rangle \Big] \;.
 \label{szbasis}
\end{equation}
This simple finding, namely the strong fragmentation of the
ground-state probability in the $S_{z}$ basis---but not in
the plaquette basis---represents one of our central results.

 The most challenging part of the calculation in the plaquette
basis is the computation of the matrix elements of the coupling
potential $V(1,2)$; recall that in the $S_z$ basis matrix elements 
of the Hamiltonian can be evaluated by inspection. Although by no
means trivial, the sort of computations which arise in this new basis 
are done routinely in atomic~\cite{consho51} and nuclear 
physics~\cite{bohmot69,shafes74}. Indeed, over the years a 
sophisticated formalism---known generically as Racah algebra---has 
been developed to tackle these computations~\cite{bieden81,brisat93}. 
It is on these techniques that we rely heavily to compute the matrix 
elements of $V(1,2)$. We find
\begin{eqnarray}
   \langle&&(l'_{1}l'_{2})j'_{1},(l'_{3}l'_{4})j'_{2},jm|V(1,2)|
          (l_{1}l_{2})j_{1},(l_{3}l_{4})j_{2},jm\rangle =
	  {3 \over 4}(-1)^{l_{2}+l_{3}+j}
	  \widehat{\jmath_{1}}\widehat{\jmath_{2}}
          \widehat{\jmath'_{1}}\widehat{\jmath'_{2}}
	  \left\{\matrix{j_{1}  & j'_{1} & 1 \cr
                         j'_{2} & j_{2}  & j \cr}\right\} 
							\nonumber \\
        \Bigg[ 
        &&\delta_{l_{2}l_{2}'}\delta_{l_{3}l_{3}'}
          \widehat{l_{1}}\widehat{l_{4}}
          \widehat{l'_{1}}\widehat{l'_{4}}
	  \left\{\matrix{l_{1}  & l'_{1} & 1     \cr
                         j'_{1} & j_{1}  & l_{2} \cr}\right\} 
	  \left\{\matrix{l_{1}  & l'_{1} & 1     \cr
                         1/2    & 1/2    & 1/2   \cr}\right\} 
	  \left\{\matrix{l_{4}  & l'_{4} & 1     \cr
                         j'_{2} & j_{2}  & l_{3} \cr}\right\} 
	  \left\{\matrix{l_{4}  & l'_{4} & 1     \cr
                         1/2    & 1/2    & 1/2   \cr}\right\} \pm 
							\nonumber \\ 
        &&\delta_{l_{1}l_{1}'}\delta_{l_{4}l_{4}'}
          \widehat{l_{2}}\widehat{l_{3}}
          \widehat{l'_{2}}\widehat{l'_{3}}
	  \left\{\matrix{l_{2}  & l'_{2} & 1     \cr
                         j'_{1} & j_{1}  & l_{1} \cr}\right\} 
	  \left\{\matrix{l_{2}  & l'_{2} & 1     \cr
                         1/2    & 1/2    & 1/2   \cr}\right\} 
	  \left\{\matrix{l_{3}  & l'_{3} & 1     \cr
                         j'_{2} & j_{2}  & l_{4} \cr}\right\} 
	  \left\{\matrix{l_{3}  & l'_{3} & 1     \cr
                         1/2    & 1/2    & 1/2   \cr}\right\} 
        \Bigg] \;.
\end{eqnarray}
Note that we have defined $\hat{x}\equiv\sqrt{2x+1}$ and that 
the $+/-$ sign in the above expression should be adopted whenever
$l_{1}+l'_{2}+l'_{3}+l_{4}+j_{1}+j'_{1}+j_{2}+j'_{2}=$~even/odd.
Aside from simple phases and numerical factors, the matrix
elements of $V(1,2)$ depend on scalar functions (i.e., independent
of $m$) known as Racah coefficients; here we cast the matrix elements 
in terms of the more symmetric $6\!-\!j$ symbols~\cite{bieden81,brisat93}.  
Closed-form expressions are readily available for the numerical 
computation of these (and many other) recoupling 
coefficients~\cite{roten59}. 

\section{Applications}
\label{sec:applic}

 In the present section we concentrate on small ($2 \times 4$ and
$2 \times 6$) ladders to illustrate some of the advantages of the 
new basis. Our aim is to provide convincing evidence, by means of 
a few simple examples, of the utility of the plaquette basis. 
Hopefully, the evidence will be strong enough to persuade some 
researchers in the field to exploit this basis in future 
calculations. 

\subsection{Distribution of ground-state strength}
\label{sec:strength}

 In Fig.~\ref{figone} we display the ground-state probabilities 
($|\langle\alpha|\Psi_{0}\rangle|^{2}$ with $\alpha$ an element 
of the basis) for the $2 \times 4$ Hamiltonian. This picture 
emphasizes what was already evident from Eq.~(\ref{szbasis}),
namely, a strong fragmentation of ground-state strength among the
many (70) states in the $S_{z}$ basis. Indeed, the only basis
states containing a nontrivial amount ($\sim 15\%$) of strength
are the (staggered) N\'eel states. In contrast, in the plaquette 
basis a single state carries 85\% of the ground-state probability
and with two states one can practically account for the full probability.
Moreover, this picture deteriorates little as one increases 
the size of the system. Indeed, for a $2 \times 6$ system one basis 
state---out of 132---carries in excess of 70\% of the ground-state 
probability and with only three states one can account for almost
90\% of it. In Fig.~\ref{figtwo} we show the corresponding distribution 
of strength for the first excited state of the system. Most of the 
features observed for the ground state remain valid in this case as 
well. Based on this evidence, we believe that the plaquette basis 
could prove very useful in numerical computations of ladder compounds.

\subsection{Contractor renormalization group method}
\label{sec:core}

 Given that most of the physics of the ground state (and of the
low-lying excited states) is contained in a very few numbers of
plaquette-basis states, it seems natural to attempt some form 
of truncation of the basis so that larger systems may be more readily
simulated
in the computer. The choice becomes obvious upon glancing at 
Table~\ref{tableone}; one should retain the first four states, having 
energies $E=-2$ and $E=-1$ (three-fold degenerate), respectively. 
Note that with this choice the link angular momenta---the high-energy 
degrees of freedom in the theory---have been ``frozen'' at 
$l_{1}=l_{2}=1$, while the plaquette angular momentum, having values of 
$j_{1}=0,1$, becomes the effective low-energy degree of freedom in
the new theory. 
In this way the original theory defined on a 16-dimensional Hilbert 
space (per plaquette) will get mapped into a new effective theory 
having the same low-energy physics as the original theory, but 
defined on a Hilbert space of one fourth the size. The  
low-energy effective theory will now be constructed using the 
newly developed COntractor REnormalization group (CORE) 
method~\cite{morwei94}.

 CORE provides a systematic approach at constructing the new
low-energy theory using contraction and cluster-expansion 
techniques. The first step into the implementation of CORE is 
the selection of an ``elementary'' block and a truncation
scheme. In our case the block is the $2 \times 2$ plaquette
and the truncation scheme has been described above; the link
variables become frozen and the plaquette variable is limited
to take the values $j_{1}=0,1$. Constructing an effective
Hamiltonian with the same low-energy properties as the original
theory---on a system that contains only one single block 
($B_{1}$)---is straightforward. We obtain,
\begin{equation}
  \langle j'_{1}m'_{1}|H_{\rm eff}(B_{1})|j_{1}m_{1}\rangle=
  \left[{1\over 2}j_{1}(j_{1}+1)-2\right]
  \delta_{j_{1}j_{1}'}\delta_{m_{1}m_{1}'} \;.
 \label{range1}
\end{equation}
This low-energy Hamiltonian constitutes the range-1 term in the
cluster expansion and is denoted by 
$h_{1}(B_{1})=H_{\rm eff}(B_{1})$. 
One now proceeds to calculate the range-2 contribution to the
cluster expansion by computing an effective Hamiltonian on a
system that contains two connected blocks ($B_{1}$ and $B_{2}$).
This is the $2 \times 4$ Hamiltonian of Eq.~(\ref{htwoplaquette}).
In constructing the various contributions to the cluster expansion
one must pay particular attention to the overlaps between the
exact eigenstates of the original Hamiltonian and the low-energy 
basis. Since it is simpler to study these overlaps in Hilbert spaces
which reflect the symmetries of the Hamiltonian, we work with a 
low-energy basis of definite total spin; recall that 
$[H,{\bf J}]=0$, where ${\bf J}\equiv{\bf S}=\sum_{i}{\bf S}_{i}$
is the total spin of the system. 

We start with the $j\!=\!0$ sector. In this sector, there are 14 
eigenstates of the exact $2\times 4$ Hamiltonian, while there are 
only two low-energy states in the truncated basis, namely, 
$|\phi_{1}\rangle\equiv|j_{1}\!=\!j_{2}\!=0,j\!=\!m\!=\!0\rangle$ 
and
$|\phi_{2}\rangle\equiv|j_{1}\!=\!j_{2}\!=1,j\!=\!m\!=\!0\rangle$. 
These two states have a nonzero overlap with the exact ground
state $|E^{(0)}_{0}\rangle$. Indeed, these are the two states 
that dominate the ground-state probability in Fig.~\ref{figone}. 
In particular, this implies that both of these basis states will 
``contract'' onto the same eigenstate of the Hamiltonian, i.e.,
\begin{equation}
  \lim_{t\rightarrow\infty} e^{-tH}|\phi_{1}\rangle 
  \propto |E^{(0)}_{0}\rangle \;\; {\rm and} \;\;
  \lim_{t\rightarrow\infty} e^{-tH}|\phi_{2}\rangle 
  \propto |E^{(0)}_{0}\rangle \;.
 \label{contractbad}
\end{equation}
CORE demands that only one low-energy basis state should
contract into the ground state. CORE also offers a simple
solution to this problem: construct a new truncated basis by 
performing a similarity transformation on the original one so that
each state in the new basis ($\xi_{i}$) contracts onto a 
unique eigenstate of the exact Hamiltonian, i.e., 
\begin{equation}
  \lim_{t\rightarrow\infty} e^{-tH}|\xi_{1}\rangle 
  \propto |E^{(0)}_{0}\rangle \;\; {\rm but} \;\;
  \lim_{t\rightarrow\infty} e^{-tH}|\xi_{2}\rangle 
  \propto |E^{(0)}_{1}\rangle \;,
 \label{contractgood}
\end{equation}
where $E^{(0)}_{1}$ is the second lowest eigenvalue 
of the Hamiltonian in the $j\!=\!0$ sector. This is all
that is needed. In this way, the relevant matrix elements 
of the effective Hamiltonian in the $j\!=\!0$ sector
become:
\begin{eqnarray}
    \langle j'_{1}j'_{2}j\!=\!0|| &&
      H_{\rm eff}(B_{1},B_{2})
    ||j_{1}j_{2}j\!=\!0\rangle = \nonumber \\ &&
    \left(\matrix{\cos\theta_{0} & -\sin\theta_{0}            \cr
                  \sin\theta_{0} &  \phantom{-}\cos\theta_{0} \cr}
    \right) 
    \left(\matrix{E^{(0)}_{0}  &  0           \cr
                   0           &  E^{(0)}_{1} \cr}
    \right) 
    \left(\matrix{\phantom{-}\cos\theta_{0} & \sin\theta_{0}  \cr
                 -\sin\theta_{0}            & \cos\theta_{0}  \cr}
    \right)\;, 
 \label{hzero}
\end{eqnarray}
where $E^{(0)}_{0}=-4.293$, $E^{(0)}_{1}=-2.500$, and
$\theta_{0}=-18.482^\circ$.

The construction of the effective Hamiltonian in the $j\!=\!1$ 
sector proceeds in a similar fashion. In this sector there are 
28 eigenstates and three low-energy basis states (each with a 
three-fold degeneracy). These are:
$|\phi_{3}\rangle\equiv|j_{1}\!=\!1\,j_{2}\!=\!0,j\!=\!1\,m\rangle$, 
$|\phi_{4}\rangle\equiv|j_{1}\!=\!0\,j_{2}\!=\!1,j\!=\!1\,m\rangle$,
and 
$|\phi_{5}\rangle\equiv|j_{1}\!=\!j_{2}\!=\!1,j\!=\!1\,m\rangle$. 
Since, in this particular case, only the first two states need to 
be transformed, the similarity transformation can 
again be parameterized in terms of a single angle. That is,
\begin{eqnarray}
    \langle j'_{1}j'_{2}j\!=\!1|| &&
      H_{\rm eff}(B_{1},B_{2})
    ||j_{1}j_{2}j\!=\!1\rangle = \nonumber \\ &&
    \left(\matrix{\cos\theta_{1} & -\sin\theta_{1}            & 0\cr
                  \sin\theta_{1} &  \phantom{-}\cos\theta_{1} & 0\cr
			0        &            0		      & 1\cr}
    \right) 
    \left(\matrix{E^{(1)}_{0}  &  0           & 0\cr
                   0           &  E^{(1)}_{1} & 0\cr
		   0           &  0           & E^{(1)}_{2}\cr}
    \right) 
    \left(\matrix{\phantom{-}\cos\theta_{1} & \sin\theta_{1}  & 0\cr
                 -\sin\theta_{1}            & \cos\theta_{1}  & 0\cr
			0        &            0		      & 1\cr}
    \right) \;, 
 \label{hone}
\end{eqnarray}
where $E^{(1)}_{0}=-3.523$, $E^{(1)}_{1}=-2.915$, $E^{(1)}_{2}=-2.590$, 
and $\theta_{1}=45^\circ$.

Finally, since (up to a five fold degeneracy) there is a unique 
$j\!=\!2$ state in the low energy basis, namely,
$|\phi_{6}\rangle\equiv|j_{1}\!=\!j_{2}\!=\!1,j\!=2\,m\rangle$, the
effective Hamiltonian in this sector is simply given by
\begin{equation}
    \langle j'_{1}j'_{2}j\!=\!2|| 
      H_{\rm eff}(B_{1},B_{2})
    ||j_{1}j_{2}j\!=\!2\rangle = E^{(2)}_{0}=-2.207 \;.
 \label{htwo}
\end{equation}
Collecting all the above results we can now write the effective
$2 \times 4$ low-energy Hamiltonian in the direct product basis:
\begin{eqnarray}
  \langle j'_{1}m'_{1}&&,j'_{2}m'_{2} 
   |H_{\rm eff}(B_{1},B_{2})|j_{1}m_{1},j_{2}m_{2}\rangle=
  \delta_{m'_{1}+m'_{2},m_{1}+m_{2}} \nonumber \\ &&
  \sum_{j}
    \langle j'_{1}j'_{2}j|| 
      H_{\rm eff}(B_{1},B_{2})
    ||j_{1}j_{2}j\rangle  
    \langle j'_{1}m'_{1},j'_{2}m'_{2}|j\,m'_{1}+m'_{2}\rangle
    \langle j_{1}m_{1},j_{2}m_{2}|j\,m_{1}+m_{2}\rangle \;,
 \label{heff2}
\end{eqnarray}  
where $\langle j_{1}m_{1},j_{2}m_{2}|jm\rangle$ are
Clebsch-Gordan coefficients. The range-2 contribution to the 
cluster expansion is obtained by simply removing from 
$H_{\rm eff}(B_{1},B_{2})$ those range-1 terms that have 
already been included in the single-block calculation, i.e.,
\begin{equation}
  h_{2}(B_{1},B_{2}) = H_{\rm eff}(B_{1},B_{2}) 
                     - h_{1}(B_{1}) - h_{1}(B_{2}) \;.
 \label{range2}
\end{equation}

To construct the renormalized Hamiltonian one must continue
this procedure, indefinitely, on larger and larger connected 
blocks. Here we will stop at the range-2 contribution. Note
that for our choice of basis, this range-2 approximation 
already takes into account correlations among next-to-next 
nearest neighbors. The approximate (up to range-2) renormalized 
Hamiltonian becomes
\begin{equation}
  H_{\rm ren}=\sum_{j=1}^{\infty}
     \Big[h_{1}(B_{j}) + h_{2}(B_{j},B_{j+1})\Big] \;.
 \label{hren}
\end{equation}
It is instructive to use this approximation to compute the
low-energy spectrum of the $2 \times 6$ (three-plaquette)
Heisenberg antiferromagnet. For this case, the range-2
approximation yields
\begin{equation}
  H_{\rm ren}=h_{1}(B_{1})+h_{1}(B_{2})+h_{1}(B_{3})
             +h_{2}(B_{1},B_{2})+h_{2}(B_{2},B_{3}) \;.
 \label{hthreeblock}
\end{equation}
This expression is useful as it suggests when the cluster 
expansion might become rapidly convergent. If an optimal 
basis has been chosen---provided that one exists---one
might hope that most of the low-energy spectrum could be 
generated by the range-1 terms, leaving the range-2 
terms in charge of the fine tuning. Alternatively, an
optimal basis could generate---dynamically---two energy 
scales in the problem; a large one associated with 
physics within the blocks and a small one associated 
with the ``residual interaction'' between the blocks.
Our investigations show that---for our truncated basis---such 
is indeed the case.  First, one can simply compare the individual 
matrix elements of $h_1$ and $h_2$.  Those of the former are 
typically 4 to 8 times larger than those of the latter.  
Second, one can arrive at the same conclusion by mapping the
original isotropic spin-$1/2$ ladder onto an equivalent 
anisotropic spin-$1$ ladder as follows. Note that for the 
2-plaquette system [see Eq.~(\ref{htwoplaquette})] 
\begin{equation}
	H_0(1) + H_0(2)=
	({\bf S_1}+{\bf S_4})\cdot({\bf S_2}+{\bf S_3})
       +({\bf S_5}+{\bf S_8})\cdot({\bf S_6}+{\bf S_7})
       ={\bf L_1}\cdot{\bf L_2} + {\bf L_3}\cdot{\bf L_4} \;.
 \label{spin11}
\end{equation}
In the truncated basis, all link angular momenta are 1; hence
the ${\bf L_i}$ are spin-$1$ operators.  Now consider 
$V(1,2)=({\bf S}_{3}\cdot{\bf S}_{5}+{\bf S}_{4}\cdot{\bf S}_{6}).$
In the truncated basis, symmetries permit the interaction to be 
written as
\begin{eqnarray}
	V(1,2)=&&{\bf S_3}\cdot{\bf S_5} + {\bf S_4}\cdot{\bf S_6}
						\nonumber \\
	\rightarrow&&{1\over 4}\bigl[
	({\bf S_2}+{\bf S_3})\cdot({\bf S_5}+{\bf S_8})
       +({\bf S_1}+{\bf S_4})\cdot({\bf S_6}+{\bf S_7}) \bigr]
						\nonumber \\
       =&&{1\over 4} 
        [{\bf L_1}\cdot{\bf L_4} + {\bf L_2}\cdot{\bf L_3} ] \ .	
 \label{spin12}
\end{eqnarray}
Hence, in the truncated basis---which carries much of the important 
physics of the problem---the isotropic 2-plaquette spin-$1/2$ system 
is equivalent to a 1-plaquette spin-$1$ system where the coupling 
along the rungs (${\bf L_1}\cdot{\bf L_2} + {\bf L_3}\cdot{\bf L_4}$) 
is four times as strong as the coupling between the rungs
(${1\over 4}
        [{\bf L_1}\cdot{\bf L_4} + {\bf L_2}\cdot{\bf L_3} ]$).
In this context our basis is optimal in the sense that the much 
stronger ``rung'' couplings are diagonal in it. We also note that
this mapping, in conjunction with CORE techniques similar to those
discussed above, is likely to provide a useful starting point for 
formulating a renormalization group transformation which can permit 
us to estimate properties of infinite ladders.  This will be the 
topic of a future publication.

In Table~\ref{tabletwo} we display the low-energy spectrum of 
the $2 \times 6$ Hamiltonian using a variety of approximations;
the ratio to the exact value appears in parenthesis. Recall 
that the spectrum has been computed with open boundary
conditions. The states have been classified according to their
total spin, which is listed in the first column. In the second 
column we report the results from an extreme ``weak-coupling'' 
calculation. In this approximation the $2 \times 2$ plaquettes 
are treated exactly but the residual interaction between the 
plaquettes is neglected [i.e., $V(j,j+1)\equiv 0$; see 
Eq.~(\ref{htwoplaquette})]. We observe that at the 80 to 90 percent 
level, the spectrum is, indeed, accounted for by the mere selection 
of the basis. In the third column we report a calculation which
uses the exact $2 \times 6$ Hamiltonian but with the truncated
low-energy basis. 
Departures from the exact results are only 5 to 10\%.
In principle, this truncation could enable
the simulations of larger systems, as the number of states 
increases with the number of plaquettes ($N_{p}$) only as 
$4^{N_{p}}$, rather than as $4^{2N_{p}}$. However, in many 
applications, this level of accuracy may still be insufficient. 
Amazingly, a dramatic improvement on these calculations results
from expending the very little additional effort required  
to construct the CORE effective Hamiltonian for the truncated 
basis. As shown in Table~\ref{tabletwo}, CORE-improved 
calculations yield results that range from a fraction of 1\%
to a few percent of the exact answer. Note that these 
results were obtained by the diagonalization of---at 
most---$20 \times 20$ matrices; instead, the exact calculation
in the $S_{z}$ basis requires a diagonalization of a 
$924 \times 924$ matrix. In the near future, we 
plan to use this renormalized Hamiltonian, perhaps including 
range-3 contributions, to simulate larger systems. 

 We conclude this section with a brief comment about the
doping of the ladders. In order to gain some qualitative
insight into the nature of hole correlations, Dagotto and
collaborators introduced two energy scales in the problem: 
a large exchange coupling $J'$ along the rungs relative
to a small exchange coupling $J$ along the chains~\cite{drs92}. 
They observed that if a pair of holes is added to the system,
the energy will get minimized whenever the two holes go
into the same rung in the ladder. Further, they concluded 
by means of numerical evidence, that most of the arguments 
developed for the anisotropic case remain valid even when 
$J'\approx J$; note that to date, the physical realization 
of the ladders seem to obey the isotropic relation~\cite{dagric96}. 
In this paper we have only considered the isotropic case. 
A particularly gratifying aspect of the mapping described above
is the natural appearance in the isotropic system of distinct 
energy scales in the effective spin-$1$ ladder. This appears
to permit carrying over the arguments of Dagotto and collaborators
with little modification.  Now it is the rungs on the spin-$1$
ladder which are strongly bound, consisting of a pair of ``frozen''
spin triplets coupled to an overall angular momentum of zero; 
interactions between rungs are relatively weak.  In terms of the
original spin-$1/2$ ladder, we may conclude that individual 
plaquettes interact only weakly. Hence the weak-coupling limit
defined above should be a reasonable approximation when applied
to very large systems just as we have found it to be by comparing 
with exact results for the 3-plaquette case. In this limit it
is simple to see that the ground-state of the system consists of 
all plaquettes being in the lowest $j\!=\!0$ state, with energy per 
plaquette of $-2$ (see Table~\ref{tableone}). When a pair of holes 
is introduced into the system the holes can go into two different 
plaquettes at a cost in energy of $+2$ (the lowest energy of 
three spins in a plaquette is equal to $-1$). Alternatively, 
the holes can go into a rung---or along a chain---in the
same plaquette at a cost in energy of $+5/4$ (the lowest 
energy of two spins in a plaquette is equal to $-3/4$). Hence, 
it becomes energetically favorable for the two holes to bind and 
break as few $j\!=\!0$ plaquettes as possible. Moreover, as 
another pair of holes is added into the system, it becomes 
energetically favorable---at least for holes with no mobility---for 
the four holes to go into the same plaquette, rather than for the 
new pair to break another plaquette. 

\section{Conclusions}
\label{sec:concl}
 
 We have employed a plaquette basis for the study of antiferromagnetic 
Heisenberg ladders. The states in the basis represent the eigenstates 
of the $2 \times 2$ plaquette and are constructed from the 
angular-momentum coupling of the four spins in the plaquette. 
Matrix elements of the Hamiltonian were computed in this basis 
and were expressed in terms of a product of five Racah coefficients. 
These expressions are considerably more complicated than the 
corresponding ones obtained using the conventional $S_{z}$ basis. 
Yet, they can be efficiently computed by employing angular-momentum 
techniques that have been developed over the years in atomic and 
nuclear physics. Moreover, this basis seems to capture some of the
important physics of these complicated systems. Indeed, we have 
shown that the distribution of ground-state and first-excited-state 
strength is concentrated in a very few number of states. This is in
contrast to the $S_{z}$ basis where the strength is strongly
fragmented. This concentration of strength among a few states provides
a very natural truncation scheme for the basis. We selected a 
low-energy basis which reduces the size of the Hilbert space by a 
factor of four per plaquette. By using the original $(2 \times 6)$ 
Hamiltonian in this truncated space we obtained a low-energy spectrum 
that was within 10\% of the exact answer. However, by improving the 
Hamiltonian---via CORE---we were able to get within 1\% of the exact 
answer. Moreover, the mere selection of the basis dynamically
generates two-energy scales in the problem---even in the case of an 
isotropic coupling. This phenomenon is most clearly understood by 
mapping the original isotropic spin-$1/2$ ladder onto an effective 
spin-$1$ ladder in which the coupling along rungs is four times 
stronger than the coupling between rungs. This separation of scales 
is important for the development of qualitative insights into the 
nature of hole doping. In particular, it supports the notion that 
holes will go into the ladders in such a way as to disturb the
minimum number of $j\!=\!0$ plaquettes.
 
 In summary, we have provided solid evidence in support of a
plaquette basis for the study of Heisenberg ladders. 
Much work remains to be done, such as the implementation of a Lanczos 
diagonalization procedure and the study of larger systems using
the renormalized range-2 (or even range-3) CORE Hamiltonian.
Yet, we believe that the mere selection of a basis could
play a prominent role in the elucidation of the important
physics behind the ladder materials.

\acknowledgments
We thank E. Dagotto for introducing one of us (JP) to some 
of the issues addressed in this work and for many useful and 
stimulating discussions. This work was supported by the DOE 
under Contracts Nos. DE-FC05-85ER250000, DE-FG05-92ER40750 
and DE-FG03-93ER40774.

\begin{figure}
 \caption{Distribution of strength for the ground state of the 
	  $2 \times 4$ Hamiltonian in the $S_{z}$ basis and in 
	  the plaquette basis.}
 \label{figone}
\end{figure}

\begin{figure}
 \caption{Distribution of strength for the first-excited state of 
	  the $2 \times 4$ Hamiltonian in the $S_{z}$ basis and in 
	  the plaquette basis.}
 \label{figtwo}
\end{figure}
\mediumtext
 \begin{table}
  \caption{Eigenvalues and eigenvectors of the $2 \times 2$ system.}
   \begin{tabular}{cccrc}
     $l_{1}$  &  $l_{2}$ & $j_{1}$ &    E  & degeneracy \\
     \tableline
        1     &     1    &    0    &  $ -2$   &   1   \\
        1     &     1    &    1    &  $ -1$   &   3   \\
        0     &     0    &    0    &  $\ 0$   &   1   \\
        0     &     1    &    1    &  $\ 0$   &   3   \\
        1     &     0    &    1    &  $\ 0$   &   3   \\
        1     &     1    &    2    &  $ +1$   &   5   \\
   \end{tabular}
  \label{tableone}
 \end{table}

\mediumtext
 \begin{table}
  \caption{Low-lying spectrum for the $2\times 6$ Heisenberg ladder.
	   The different approximations are explained in the text.	   
           Quantities in parenthesis represent the ratio to the
	   exact value.}
   \begin{tabular}{ccccc}
   $S$ & Weak Coupling & Truncated & CORE (Range-2) & Exact \\
     \tableline
 0 & $-6.000~(0.909)$ & $-6.335~(0.959)$ & $-6.588~(0.998)$ & $-6.603$ \\
   & $-4.000~(0.818)$ & $-4.396~(0.899)$ & $-4.899~(1.002)$ & $-4.888$ \\
   & $-4.000~(0.836)$ & $-4.250~(0.889)$ & $-4.882~(1.021)$ & $-4.783$ \\
     \tableline
 1 & $-5.000~(0.844)$ & $-5.580~(0.942)$ & $-5.928~(1.001)$ & $-5.924$ \\
   & $-5.000~(0.909)$ & $-5.112~(0.930)$ & $-5.458~(0.993)$ & $-5.498$ \\
   & $-5.000~(0.983)$ & $-4.666~(0.917)$ & $-5.132~(1.009)$ & $-5.087$ \\
   & $-4.000~(0.797)$ & $-4.538~(0.905)$ & $-5.092~(1.015)$ & $-5.017$ \\
   & $-4.000~(0.850)$ & $-4.125~(0.877)$ & $-4.809~(1.022)$ & $-4.705$ \\
     \tableline
 2 & $-4.000~(0.823)$ & $-4.413~(0.908)$ & $-4.862~(1.000)$ & $-4.862$ \\
   & $-4.000~(0.925)$ & $-3.875~(0.890)$ & $-4.426~(1.017)$ & $-4.352$ \\
     \tableline
 3 & $-3.000~(0.895)$ & $-2.750~(0.821)$ & $-3.414~(1.019)$ & $-3.351$ \\
   \end{tabular}
  \label{tabletwo}
 \end{table}

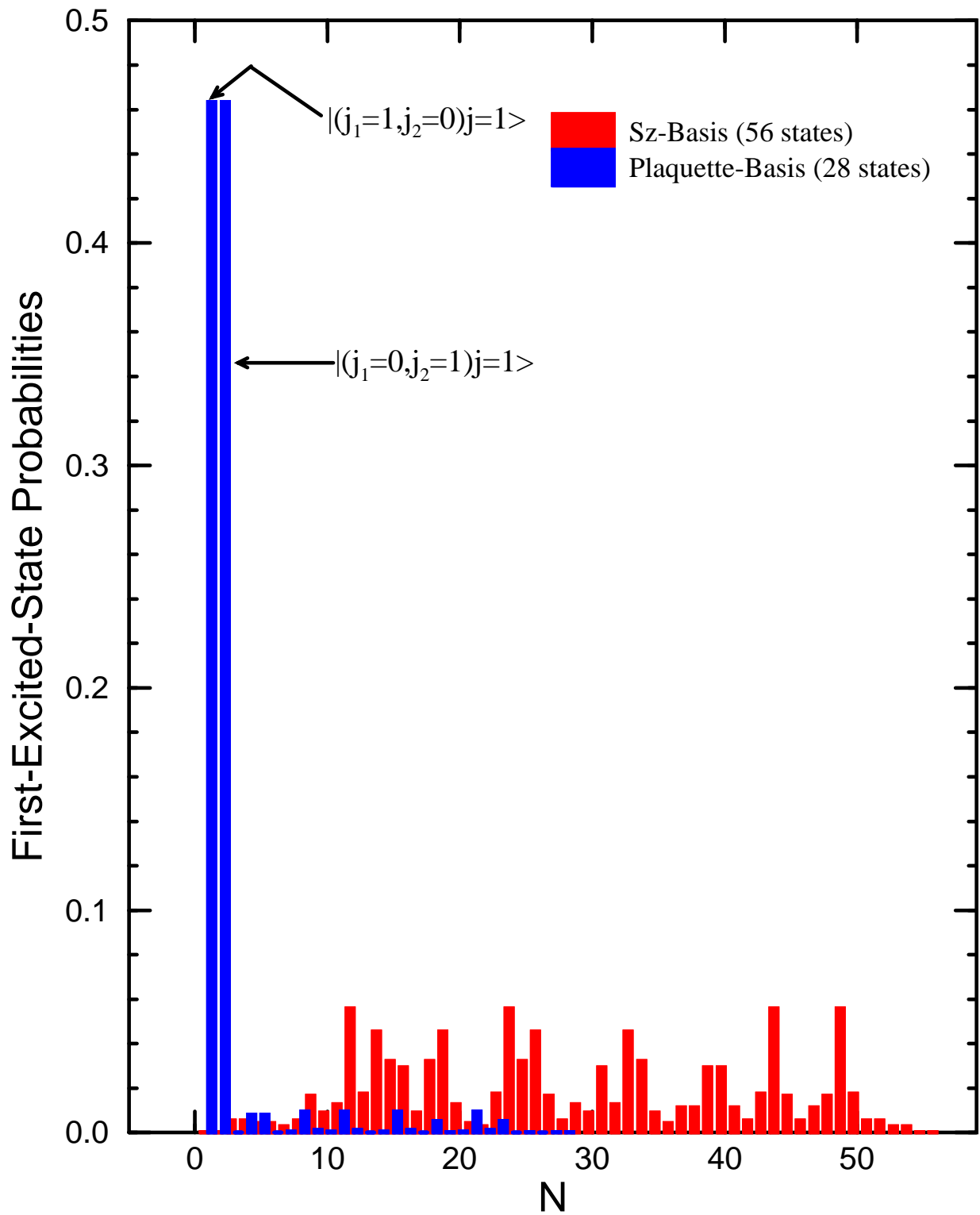
\end{document}